\documentclass[pra,twocolumn,tightenlines,showpacs,nofootinbib]{revtex4}
\usepackage{multirow}
\usepackage{amsmath}
\usepackage{bm}
\usepackage{dcolumn}

\begin{document}
\title{Anomalously small blackbody radiation shift in Tl$^+$ frequency standard}

\author{Z. Zuhrianda$^{1}$}
\author{M. S. Safronova$^1$}
\author{M. G. Kozlov$^2$}

\affiliation{$^1$ Department of Physics and Astronomy, University of Delaware,
                  Newark, Delaware 19716, USA}
\affiliation{$^2$ Petersburg Nuclear Physics Institute, Gatchina,
                  Leningrad District, 188300, Russia}

\date{\today}

\begin{abstract}
The operation of atomic clocks is generally carried out at room temperature, whereas the definition of the second refers to the clock transition in
an atom at absolute zero. This implies that the clock transition frequency should be corrected in practice for the effect of finite temperature of
which the leading contributor is the blackbody radiation (BBR) shift.    In the present work, we  used configuration interaction + coupled-cluster
method  to evaluate polarizabilities of the $6s^2~^1S_0$ and $6s6p~^3P_0$ states of Tl$^+$ ion; we find $\alpha_0(^1S_0)=19.6$~a.u. and
$\alpha_0(^3P_0)=21.4$~a.u.. The resulting BBR shift of the $6s6p~^3P_0 - 6s^2~^1S_0$ Tl$^+$ transition at $300~K$ is $\Delta \nu_{\rm
BBR}=-0.0157(16)$~Hz. This result demonstrates that  near cancelation of the $^1S_0$ and $^3P_0$ state polarizabilities  in divalent B$^+$, Al$^+$,
In$^+$ ions of group IIIB [Safronova \textit{et al.}, PRL 107, 143006 (2011)] continues for much heavier Tl$^+$, leading to anomalously small BBR
shift for this system. This calculation demonstrates that the BBR contribution to the fractional frequency uncertainty of the Tl$^+$ frequency
standard at $300~K$  is $1\times10^{-18}$. We find that Tl$^+$ has the smallest fractional BBR shift among all present or proposed frequency
standards with the exception of Al$^+$.
\end{abstract}
\maketitle

\section{Introduction}

Recent advances in atomic and optical physics have led to unprecedented improvements in the accuracy of optical frequency standards that are
essential for many applications including  measurements of the fundamental constants and search of their variation with time, testing of physics
postulates, inertial navigation, magnetometry,  tracking of deep-space probes, and others \cite{KarPei10}. An optical clock with a record low
fractional frequency uncertainty of $8.6\times10^{-18}$, based on quantum logic spectroscopy
 of an Al$^+$ ion was demonstrated  in 2010 \cite{ChoHumKoe10}.

 Any definition of the second should be based on a clock decoupled from its particular environment.
    Thermal fluctuations of the electromagnetic field, i.e.  blackbody radiation (BBR), are pervasive and
     can only be suppressed by cooling the clock.  The BBR at any non-zero
     temperature induces small shifts in atomic energy levels through the AC Stark effect.
 The operation of atomic clocks is generally carried out at room temperature
     and  the clock transition frequency should be corrected in practice for the BBR  shift.
 Experimental measurements of the BBR shifts are sufficiently difficult that no direct
measurement has yet been reported for optical frequency standards. At room temperature, the BBR shift of a clock transition turns out to make one of
the largest irreducible contributions to the uncertainty budget
      of optical atomic clocks \cite{SafJiaAro10}.
 The present
  status of the
   theoretical and experimental determinations of the BBR shifts in all frequency standards was
   recently reviewed in
\cite{SafJiaAro10,MitSafCla10}.

The BBR frequency shift of a
  clock transition can be related to the difference of the static electric-dipole
  polarizabilities between the  two clock states
\cite{PorDer06}. Recent work \cite{SafKozCla11} demonstrated that the polarizabilities of ground $ns^2~^1S_0$  and metastable $nsnp~^3P_0$ states are
nearly equal to each other in  B$^+$, Al$^+$, and In$^+$, all of which are group IIIB ions. As a result, these three ions have anomalously small BBR
shifts of the $ns^2 ~^1S_0 - nsnp~ ^3P_0$ clock transitions. The fractional BBR shifts for these ions are at least 10 times smaller than those of any
other present or proposed optical frequency standards at the same temperature, and are less than 0.3\% of the Sr clock shift.

Optical frequency standard based on $^{204}$Tl$^+$ $6s^2~^1S_0~m_F=0$~--~$6s6p~^3P_0~m_{F^{\prime}}=0$ transition was proposed in
Ref.~\cite{DemYuNag89}. The radioactive isotope of $^{204}$Tl has a half-life of 3.78 years, a spin of 2, and a very small magnetic moment of 0.0908
nuclear magnetons making it ideal object for very high-resolution laser spectroscopy \cite{DemYuNag89}.  Because of its small nuclear magnetic moment
the natural linewidth of the clock transition in $^{204}$Tl$^+$ is expected to be orders of magnitude smaller than estimated for stable Tl isotopes
\cite{DemYuNag89}. The BBR in this frequency standard have not been previously estimated. Since  three group IIIB ions exhibit very small BBR shifts,
it is very interesting to evaluate if this trend holds for much heavier Tl$^+$.

The BBR frequency shift of the clock  transition  can be related to the difference of the static electric-dipole polarizabilities between the clock
states, $\Delta\alpha_0$,  by \cite{PorDer06}
\begin{equation}
 \Delta \nu_{\rm BBR} = -\frac{1}{2}(831.9~\mathrm{V/m})^2
\left( \frac{T(K)}{300} \right)^4 \Delta \alpha_0(1+\eta), \label{eq2}
\end{equation}
where  $\eta$ is a small dynamic correction due to the frequency distribution
 and only the electric-dipole transition part of the contribution is
considered.  The M1 and E2 contributions have been estimated for Al$^+$ and found to be negligible \cite{SafKozCla11}. Therefore, the calculation of
the BBR shift reduces to accurate calculation of the static polarizabilities of the clock states and dynamic correction $\eta$.

 In this work, we
evaluate polarizabilities of the $6s^2 ~^1S_0$ and   $6s6p~^3P_0$ states in Tl$^+$, corresponding BBR shift and its uncertainty. Dynamic correction
to the BBR shift is evaluated and found negligible. We also calculate a number of electric-dipole matrix elements in Tl$^+$ for transitions between
low-lying levels. We note that our calculation of all of these properties is independent on the particular isotope number well within the quoted level of
precision. Therefore, all these results apply to any Tl$^+$ isotope.

\section{Method}

Correlation corrections between a few valence electrons can be accurately treated by the configuration interaction (CI) method. Since the
valence-valence correlations are very large, the CI method provides better description of these correlations than the perturbative approaches.
However, excitations of the core $[1s^2, ..., 5d^{10}]$ electrons can not be directly included in the CI approach due to enormous size of such problem.
An elegant approach to the inclusion of the core-valence correlations within the CI framework was developed in \cite{DzuFlaKoz96b}, where
core-valence correlations were incorporated into the CI by constructing an effective Hamiltonian using the second-order many-body perturbation theory
(CI+MBPT).
 Recently, we have developed the relativistic  CI+all-order method \cite{SafKozJoh09}
combining CI with coupled-cluster (CC) approach. This method, first suggested in \cite{Koz04}, was successfully applied to the calculation of
divalent atom properties in Refs.~\cite{SafKozJoh09,SafKozCla11}. The coupled-cluster method used here is known to describe the core-core and
core-valence correlations very well as demonstrated by its great success in predicting alkali-metal atom properties \cite{SafJoh08}. Therefore,
combination of the CI and all-order coupled-cluster methods allows to account for all dominant correlations to all orders. To evaluate uncertainty of
our results, we use all three of the approaches and compare the results of the CI, CI+MBPT, and CI+all-order calculations. We refer the reader to
Refs.~\cite{DzuFlaKoz96b,SafKozJoh09,SafKozCla11} for the description of the methods and outline only main points of the calculations below.

 We start with solving Dirac-Fock  (DF) equations
 $$ \hat H_0\, \psi_c = \varepsilon_c \,\psi_c, $$
  where
$H_0$ is the relativistic DF Hamiltonian \cite{DzuFlaKoz96b,SafKozJoh09} and $\psi_c$ and $\varepsilon_c$ are single-electron wave functions and
energies. The self-consistent calculations were performed for the [$1s^2,...,5d^{10}$] closed core and the $6s$, $7s$, $6p$, $7p$, and $6d$ orbitals.
We have constructed the B-spline basis set  consisting of $N=35$ orbitals for each of the $s,~ p_{1/2},~ p_{3/2},~ ...$ partial waves up to $l\leq5$;
core, $6s$, $7s$, $6p$, $7p$, and $6d$ orbitals were replaced by the exact DF functions for increased accuracy. The basis set is formed in a
spherical cavity with radius 60 a.u.
 The CI space is effectively complete and  includes $20sp$ and $21dfg$ orbitals.  All MBPT and all-order terms were summed over the entire $N=35,~ l\leq5$ basis set.

The multiparticle relativistic equation for three valence electrons is solved within the CI framework \cite{KotTup87} to find the wave functions and
the low-lying energy levels: $$ H_{\rm eff}(E_n) \Phi_n = E_n \Phi_n.$$
 The effective Hamiltonian is defined as $$ H_{\rm eff}(E) = H_{\rm FC} +
\Sigma(E),$$ where $H_{\rm FC}$ is the Hamiltonian in the frozen-core approximation and the energy-dependent operator $\Sigma(E)$ takes into account
virtual core excitations.   The  $\Sigma(E)$ part of the effective Hamiltonian is constructed using the second-order perturbation theory in the
CI+MBPT approach \cite{DzuFlaKoz96b} and linearized coupled-cluster single-double  method  in the CI+all-order approach \cite{SafKozJoh09}. The
$\Sigma(E)=0$ in the pure CI calculation. Construction of the effective Hamiltonian in CI+MBPT and CI+all-order approximations is described in detail
in Refs.~\cite{DzuFlaKoz96b,SafKozJoh09}.

\section{Results}
Comparison of the energy levels (in cm$^{-1}$) obtained in the CI, CI+MBPT, and CI+all-order approximations with experimental values
\cite{Moo71,SanMarYou05} is given in Table~\ref{table1}.
Corresponding relative differences of these three calculations with experiment are given in the last three columns in \%. Two-electron binding
energies are given in the first row of Table~\ref{table1}, energies in other rows are counted from the ground state. We also observed significant, by
a factor of 4 or better, improvement in the precision of the energy levels with CI+all-order method in comparison with the CI+MBPT one. For example,
CI+MBPT value for the two-electron binding energy differs from experiment by 1.8\%, while our all-order value differs from the experiment by only
0.4\% (see line one of Table~\ref{table1}). The experimental value of the two-electron binding energy is obtained as the sum of the Tl$^{+}$ and
Tl$^{2+}$ ionization limits given in \cite{Moo71}, 164765(5)~cm$^{-1}$ and 240600~cm$^{-1}$. Ref.~\cite{Moo71} notes that the ionization limit for
Tl$^{2+}$, derived from the first 3 members of the $^2S$ series was shifted by 300~cm~$^{-1}$ to give effective quantum number for $5g~^2G$ that is
nearly hydrogenic. Therefore, there is some uncertainty ($\lesssim 0.1\%$) associated with the two-electron binding energy in Tl$^+$.

We also compared the transition energies between the $6s6p~^3P_0$ level and 4 levels relevant to the calculation of the $6s6p~^3P_0$ polarizability.
These values, calculated in the CI+all-order approximation are compared with  experiment in Table~\ref{table1a}.  We find that these transition
energies are substantially more accurate than the energy levels counted from the ground state listed in Table~\ref{table1}.

In the present calculation, the Tl$^{+}$ scalar polarizability $\alpha_0$
 is separated into a valence polarizability $\alpha_0^v$, ionic core polarizability $\alpha_c$, and a small term $\alpha_{vc}$  (
 that modifies ionic core polarizability due to
 the presence of two valence
electrons.
 The ionic core polarizability is evaluated in the  random-phase approximation (RPA), an approach that is expected to provide core values
accurate to better than 5\%~\cite{MitSafCla10}. We approximate the vc term by adding vc contributions from the individual electrons, i.e.
$\alpha_{vc}(6s^2)=2\alpha_{vc}(6s)$, and $\alpha_{vc}(6s6p)=\alpha_{vc}(6s)+\alpha_{vc}(6p)$.
 For consistency, this term
is also calculated in RPA. We note that $\alpha_{vc}$ contributions are small, but their contribution to the $\Delta\alpha(^3P_0-~^1S_0)$
polarizability difference is significant, 15\%, due to severe cancelation of the valence polarizabilities of these two states.
\begin{table}
\caption{\label{table1}Comparison between experimental \cite{Moo71,SanMarYou05} and theoretical energy levels in cm$^{-1}$. Two-electron binding
energies are given in the first row, energies in other rows are counted from the ground state. Results of the CI, CI+MBPT, and CI+all-order
calculations are given in columns labeled CI, MBPT, and All. Corresponding relative differences of these three calculations with experiment are given
in the last three columns in \%.}
\begin{ruledtabular}
\begin{tabular}{lrrrrrcr}
\multicolumn{1}{c}{\multirow{2}{*}{State}} & \multicolumn{1}{c}{\multirow{2}{*}{Expt.}} & \multicolumn{1}{c}{\multirow{2}{*}{CI}} &
\multicolumn{1}{c}{\multirow{2}{*}{MBPT}} &
 \multicolumn{1}{c}{\multirow{2}{*}{All}}
 &\multicolumn{3}{c}{Differences (\%)} \\
&&&&& \multicolumn{1}{c}{CI} & \multicolumn{1}{c}{MBPT} & \multicolumn{1}{c}{All} \\ \hline
$6s^2\; ^1S_0$ & 405365 & 376102 & 412676 & 407125 &  $-$7 & 1.8 & 0.4   \\
$6s7s\; ^3S_1$ & 105229 & 92945 & 108031  & 106028  & $-$12 & 2.7 & 0.8  \\
$6s7s\; ^1S_0$ & 108000 & 96304 & 110845  & 108904  & $-$11 & 2.6 & 0.8  \\
$6s6d\; ^1D_2$ & 115166 & 101238 & 118678 & 116194 &  $-$12 & 3.1 & 0.9  \\
$6s6d\; ^3D_1$ & 116152 & 103334 & 119000 & 116857 &  $-$11 & 2.5 & 0.6  \\
$6s6d\; ^3D_2$ & 116436 & 103555 & 119339 & 117284 &  $-$11 & 2.5 & 0.7  \\
$6s6d\; ^3D_3$ & 116831 & 103911 & 119688 & 117758 &  $-$11 & 2.5 & 0.8  \\
$6p^2\; ^3P_0$ & 117408 & 108495 & 120875 & 118450 &  $-$8 & 3.0 & 0.9   \\
$6p^2\; ^3P_1$ & 125338 & 114961 & 129401 & 126440 &  $-$8 & 3.2 & 0.9   \\
$6p^2\; ^3P_2$ & 128817 & 117721 & 132754 & 129839 &  $-$9 & 3.1 & 0.8   \\
$6s8s\; ^3S_1$ & 133568 & 120147 & 136369 & 134187 &  $-$10 & 2.1 & 0.5  \\
$6s8s\; ^1S_0$ & 134292 & 121089 & 137132 & 134950 &  $-$10 & 2.1 & 0.5   \\
\\
$6s6p\; ^3P_0$ & 49451  & 41719 & 52320   & 50288    &$-$16 & 5.8 & 1.7   \\
$6s6p\; ^3P_1$ & 52394  & 44743 & 55114   & 53060    &$-$15 & 5.2 & 1.3   \\
$6s6p\; ^3P_2$ & 61728  & 61728 & 65044   & 62669    &$-$14 & 5.4 & 1.5   \\
$6s6p\; ^1P_1$ & 75663  & 75663 & 76866   & 76145    &$-$7 & 1.6 & 0.6    \\
$6s7p\; ^3P_0$ & 119361 & 119361 & 122299 & 120155 & $-$11 & 2.5 & 0.7  \\
$6s7p\; ^3P_1$ & 119576 & 119576 & 122602 & 120472 & $-$11 & 2.5 & 0.8  \\
$6s7p\; ^3P_2$ & 122209 & 122029 & 124873 & 122675 & $-$11 & 2.3 & 0.5  \\
$6s7p\; ^1P_1$ & 122379 & 122379 & 126014 & 124019 & $-$9 & 3.0 & 1.3   \\
$6s5f\; ^3F_2$ & 136216 & 136216 & 138873 & 136600 & $-$10 & 2.0 & 0.3  \\
$6s5f\; ^3F_3$ & 136115 & 136115 & 138868 & 136577 & $-$10 & 2.0 & 0.4  \\
$6s5f\; ^3F_4$ & 136230 & 136230 & 138870 & 136595 & $-$10 & 1.9 & 0.3  \\
$6s5f\; ^1F_3$ & 136263 & 136263 & 138997 & 136756 & $-$10 & 2.0 & 0.4 \\
\end{tabular}
\end{ruledtabular}
\end{table}
\begin{table}
\caption{\label{table1a}Comparison between experimental \cite{Moo71,SanMarYou05} and CI+all-order transition energies in cm$^{-1}$. The relative
differences are given  in the last column in percent.}
\begin{ruledtabular}
\begin{tabular}{lrcr}
\multicolumn{1}{c} {Transition} & \multicolumn{1}{c}{Expt.} &  \multicolumn{1}{c}{CI+all-order} &
 \multicolumn{1}{c}{Dif. (\%)}
\\ \hline
  $6s6p\; ^3P_0 - 6s7s\; ^3S_1$ &  55778   &   55739   &   0.07\%    \\
  $6s6p\; ^3P_0 - 6s6d\; ^3D_1$ &  66701   &   66569   &   0.20\%    \\
  $6s6p\; ^3P_0 - 6p^2\; ^3P_1$ &  75887   &   76152   &   $-$0.35\%   \\
  $6s6p\; ^3P_0 - 6s8s\; ^3S_1$ &  84117   &   83899   &   0.26\%    \\
\end{tabular}
\end{ruledtabular}
\end{table}
\begin{table*}
\caption{\label{table2}Contributions to the  $6s^2\;^1S_0$ and $6s6p\;^3P_0$ polarizabilities in a.u. The dominant contributions to the valence
polarizabilities are listed separately with the corresponding E1 matrix elements given in columns labeled $D$. The remaining valence contribution is
given in row Other. The contribution from the core and vc terms are given by $\alpha_c$ and $\alpha_{vc}$, respectively. The dominant contributions
to $\alpha_0$ listed in columns CI+all$^A$ and CI+all$^B$ are calculated with CI + all-order energies and NIST \cite{Moo71,SanMarYou05} energies,
respectively. The differences of the $^3P_0$ and $^1S_0$ polarizabilities calculated in different approximations are given in the last row.}
\begin{ruledtabular}
\begin{tabular}{llcrcrcrr}
\multirow{2}{*}{State} & \multirow{2}{*}{Contribution} & \multicolumn{2}{c}{CI} & \multicolumn{2}{c}{CI+MBPT} & \multicolumn{2}{c}{CI+all$^A$} &
\multicolumn{1}{c}{CI+all$^\mathrm{B}$}\\[0.5pc] && $D$ & $\alpha_0~~$ & $D$ & $\alpha_0~~$ & $D$ & $\alpha_0~~$ & $\alpha_0~~$\\ \hline
\\
$6s^2\; ^1S_0$ & $6s^2\; ^1S_0 - 6s6p\; ^3P_1$ & 0.424 & 0.589 & 0.658 & 1.149 & 0.597 & 0.984 & 0.997\\
& $6s^2\; ^1S_0 - 6s6p\; ^1P_1$ & 2.789 & 16.131 & 2.619 & 13.057 & 2.646 & 13.450 &  13.535\\
& Other && 0.269 && 0.143 && 0.155 & 0.155\\
& $\alpha_c$ && 4.983 && 4.983 && 4.983 &4.983\\
& $\alpha_{vc}$ && $-$0.071 && $-$0.071 && $-$0.071 & $-$0.071\\
& Total && 21.901 && 19.261 && 19.501 & 19.599\\
\\
$6s6p\; ^3P_0$ & $6s6p\; ^3P_0 - 6s7s\; ^3S_1$ & 1.044 & 3.113 & 0.975 & 2.499 & 0.980 & 2.519 &  2.517\\
& $6s6p\; ^3P_0 - 6s6d\; ^3D_1$ & 2.007 & 9.563 & 1.893 & 7.860 & 1.897 & 7.912 &  7.897\\
& $6s6p\; ^3P_0 - 6p^2\; ^3P_1$ & 1.616 & 5.219 & 1.557 & 4.603 & 1.562 & 4.690 & 4.706\\
& Other && 1.782 && 1.630 && 1.660 & 1.660\\
& $\alpha_c$ && 4.983 && 4.983 && 4.983 & 4.983\\
& $\alpha_{vc}$ && $-$0.338 && $-$0.338  && $-$0.338 & $-$0.338\\
& Total && 24.322 && 21.236 && 21.426 & 21.425\\ \hline
\\
{$\Delta\alpha_0(^3P_0 -~ ^1S_0)$}& && 2.421 & & 1.975 && 1.925 & 1.826 \\
\end{tabular}
\end{ruledtabular}
\end{table*}
 The valence  polarizability is determined  by solving the inhomogeneous equation of perturbation theory in the valence space, which
is approximated as
\begin{equation}
\label{eq1} (E_v - H_{\textrm{eff}})|\Psi(v,M^{\prime})\rangle = D_{\mathrm{eff},q} |\Psi_0(v,J,M)\rangle
\end{equation}
for a state  $v$ with the total angular momentum $J$ and projection $M$ \cite{KozPor99a}. The wave function $\Psi(v,M^{\prime})$, where
$M^{\prime}=M+q$, is composed of parts that have angular momenta of $J^{\prime}=J,J \pm 1$ from which the scalar and tensor polarizability
 of the state $|v,J,M\rangle$ can be determined \cite{KozPor99a}.
The effective dipole operator $D_{\textrm{eff}}$ includes RPA corrections.

Unless stated otherwise, we use atomic units (a.u.) for all matrix elements and polarizabilities throughout this paper: the numerical values of the
elementary
 charge, $e$, the reduced Planck constant, $\hbar = h/2
\pi$, and the electron mass, $m_e$, are set equal to 1. The atomic unit for polarizability can be converted to SI units via
$\alpha/h$~[Hz/(V/m)$^2$]=2.48832$\times10^{-8}\alpha$~(a.u.). The conversion coefficient is $4\pi \epsilon_0 a^3_0/h$ in SI units and the Planck
constant $h$ is factored out in order to provide direct conversion into frequency units; $a_0$ is the Bohr radius and $\epsilon_0$ is the electric
constant.
\begin{table}
\caption{\label{table4}Contributions to dynamic corrections $\eta$ for $6s^2\; ^1S_0$ and $6s6p\; ^3P_0$ states.}
\begin{ruledtabular}
\begin{tabular}{llrr}
State & Transition & \multicolumn{1}{c}{$y_{n}$} &\multicolumn{1}{c}{$\eta$}\\
\hline \\
$\eta(6s^2\; ^1S_0)$ & $6s^2\; ^1S_0 - 6s6p\; ^3P_1$ &    363  &  0.000099 \\
& $6s^2\; ^1S_0 - 6s6p\; ^1P_1$                &    251  &  0.000015 \\
 && & 0.000114\\[0.5pc]
$\eta(6s6p\; ^3P_0)$ & $6s6p\; ^3P_0 - 6s7s\; ^3S_1$ &    268  &  0.000031 \\
& $6s6p\; ^3P_0 - 6s6d\; ^3D_1$                &    320  &  0.000068 \\
& $6s6p\; ^3P_0 - 6p^2\; ^3P_1$                &    364  &  0.000031 \\
 & && 0.000130\\   [0.5pc]
$\Delta \eta(^3P_0 -~ ^1S_0)$ &&&0.000016 \\
\end{tabular}
\end{ruledtabular}
\end{table}
\begin{table*}[ht]
\caption[]{ BBR shifts at $T=300K$ in   B$^+$, Al$^+$, In$^+$, and Tl$^+$. B$^+$, Al$^+$, and In$^+$ values are taken from Ref.~\cite{SafKozCla11}.
 Polarizabilities $\alpha_0$ and their differences $\Delta\alpha_0$ are given  in a.u.; clock frequencies $   \nu_0$ and the BBR
shifts $|\Delta\nu_{\textrm{BBR}}|$  are given in Hz. Uncertainties in the values of $\Delta\nu_{\textrm{BBR}}/\nu_0$ are given in column labeled
``Uncertainty''.} \label{tab4}
\begin{ruledtabular}
\begin{tabular}{lddddrrr}
\multicolumn{1}{c}{Ion} &\multicolumn{1}{c}{$\alpha_0(^1S_0)$} &\multicolumn{1}{c}{$\alpha_0(^3P_0)$}& \multicolumn{1}{c}{$\Delta \alpha_0$}
&\multicolumn{1}{r}{$\Delta\nu_{\textrm{BBR}}$ (Hz) } &\multicolumn{1}{c}{$   \nu_0$ (Hz)   } &\multicolumn{1}{c}{$ |\Delta\nu_{\textrm{BBR}}/\nu_0|
$}
&\multicolumn{1}{c}{Uncertainty }      \\
\hline
B$^+$    &9.624 & 7.772 &-1.85(19) &  0.0159(16) &  $1.119\times10^{15}  $  &  $   1.42\times10^{-17}  $   &   $   1\times10^{-18} $   \\
Al$^+$   &24.048& 24.543& 0.495(50)& -0.00426(43)&  $1.121\times10^{15}  $  &  $   3.8\times10^{-18}  $   &   $   4\times10^{-19} $   \\
In$^+$   &24.01 & 26.02 & 2.01(20) & -0.0173(17) &  $1.267\times10^{15}  $  &  $   1.36\times10^{-17} $   &   $   1\times10^{-18} $   \\
Tl$^+$   &19.60 & 21.43 & 1.83(18) & -0.0157(16) &  $1.483\times10^{15}  $  &  $   1.06\times10^{-17} $   &   $   1\times10^{-18} $   \\
    \end{tabular}
\end{ruledtabular}
\end{table*}

While we do not use the sum-over-state approach in the calculation of the polarizabilities,
it is useful to establish which
levels give the dominant contributions. We evaluate several
leading contributions to polarizabilities
 by combining our values of the E1 matrix elements and energies  according to the sum-over-states formula for the valence polarizability~\cite{MitSafCla10}:
\begin{equation}\label{genpol}
\alpha_0^v = \frac{2}{3(2J+1)}\sum_n\frac{|\langle v\| D\| n\rangle|^2}{E_n-E_v}
\end{equation}
where $J$ is the total angular momentum of state $v$, $D$ is the electric dipole operator, and $E_i$ is the energy of the state $i$.

The breakdown of the contributions to the $6s^2\ ^1S_0$ and $6s6p\ ^3P_0$ polarizabilities  $\alpha _0$ of Tl$^+$ in a.u. is given in
Table~\ref{table2}.  Absolute values of the corresponding reduced electric-dipole matrix elements are listed in column labeled ``$D$'' in $a_{0}e$.
To demonstrate the size of the correlation corrections, we list valence results obtained in the CI, CI+MBPT, and CI+all-order approximations. The
contribution of the other terms listed in the row ``Other'' is obtained by subtracting the sum of the contributions that are calculated separately
from the total valence polarizability result obtained by the direct solution of the Eq.~(\ref{eq1}). With the exception of the last column labeled
CI+all$^B$, we use the theoretical energies obtained in the respective approximations. To obtain data listed in the last column, we combine
CI+all-order E1 matrix elements and experimental energies. The polarizability of the ground state changes by 0.5\% as expected from the accuracy of
the $6s^2~^1S_0 - 6s6p~^1P_1$ transition energy listed in Table~\ref{table1}. The polarizability of the excited $6s6p~^3P_0$ state remains the same
to four significant figures. Such remarkable agreement is due to excellent, 0.07\%, accuracy of the CI+all-order $6s6p\; ^3P_0 - 6s7s\; ^3S_1$
transition energy and opposite signs of the difference between CI+all-order $6s6p\; ^3P_0 - 6s6d\; ^3D_1$ and $6s6p\; ^3P_0 - 6p^2\; ^3P_1$
transition energies and experiment (see Table~\ref{table1a}). We note that while the change in the ground state polarizability is only 0.5\%, the
corresponding change in the final polarizability difference $\Delta \alpha (^3P_0-^1S_0)$  is 5\%.

We have also calculated the dynamic correction  $\eta$ of both clock states. The total dynamic correction $\eta$ in Eq.~(\ref{eq2}) is the difference
of individual corrections, $\Delta \eta(^3P_0 - ^1S_0)=\eta(^3P_0)-\eta(^1S_0)$. The dynamic correction $\eta$  of the   state $v$ is evaluated using
the formula~\cite{PorDer06}
\begin{equation}
\eta = \sum_n\frac{(80/63)\pi^2}{\alpha_0 T}\frac{|\langle
v\|D\|n\rangle|^2}{(2J+1)y_n^3}\left(1+\frac{21\pi^2}{5y_n^2}+\frac{336\pi^4}{11y_n^4}\right),\nonumber
\end{equation}
where $y_n=\omega_{nv}/T$; $\alpha_0$ is the static dipole polarizability of the state $v$, and $J$ the  total angular
 momentum of the state $v$. We list the dominant contributions to $\eta$ of the clock states calculated using CI+all-order E1 matrix elements and experimental energies
 in Table \ref{table4}. The sum in the expression for $\eta$ above converges very rapidly making all other contributions negligible.
 The values of $\eta$  for the $6s\;^1S_0$ and $6s6p\;^3P_0$ state are almost equal, and their
 difference listed in the last row gives only 0.0016\% contribution to the BBR shift.

\section{Evaluation of the uncertainty and conclusion}

We use Table ~\ref{table2} to evaluate the uncertainty to the BBR shift due to the core-valence correlation corrections by comparing the CI, CI+MBPT,
  and CI+all-order results for $\Delta\alpha_0(^3P_0 - ^1S_0)$ listed in the last row of Table~\ref{table2}.
The difference between the CI and CI+MBPT results is 23\%, which is expected owing to poor agreement of CI energies with experiment. The difference
between the CI+MBPT and CI+all-order results is only 3\%. As we noted above, the use of the experimental energies changes CI+all-order value by 5\%.

We studied the effect of the Breit interaction by repeating the CI+all-order calculation with the one-body part of the Breit interaction incorporated
into the DF equations and construction of the basis set on the same footing with the Coulomb interaction. We find that the Breit interaction affects
both $^1S_0$ and $^3P_0$ polarizabilities by approximately the same amount, $-0.5\%$. As a result, the correction to the BBR shift due to the Breit
interaction  is negligible (0.6\%)  at the present level of accuracy.

To evaluate the uncertainty in the $\alpha_{vc}$ contribution to the polarizability, we calculate this term in both DF and RPA approximations.  The
difference between these results is taken to be the uncertainty. We find that the uncertainty of the vc term contributes 2.4\% to the uncertainty in
the BBR shift. The ionic core polarizability $\alpha_{c}$ is the same for both states and does not contribute to the BBR shift.

Based on the comparison of the CI, CI+MBPT, and CI+all-order data, estimated of the accuracy of the $\alpha_{vc}$ terms, and estimated effect of the
Breit interaction, we place an upper bound on the uncertainty of our $\Delta\alpha_0(^3P_0 - ^1S_0)$ polarizability difference and the corresponding
BBR shift of Tl$^+$
 at 10\%.

Our final result for the BBR shift of the $6s^2~^1S_0-6s6p~^3P_0$ transition in Tl$^+$ is $\Delta \nu_{\rm BBR}=-0.0157(16)$~Hz at $300~K$.
 The corresponding relative BBR shift at $300~K$ is $|\Delta \nu_{\rm BBR}/\nu_0|=1.1(1)\times10^{-17}$.
Our final results are summarized in Table~\ref{tab4}, where we list the clock state polarizabilities, their
 difference $\Delta\alpha_0$,  BBR
shift at $T=300~$K, $^1S_0-^3P_0$ clock frequencies $\nu_0$,
 absolute value of the relative BBR shift $|\Delta\nu_{\textrm{BBR}}/\nu_0|$, and the uncertainty in the relative
 BBR shift of Tl$^+$. The Tl$^+$ values are compared with the results obtained for B$^+$,
Al$^+$, and In$^+$ ions in Ref.~\cite{SafKozCla11}.   The results listed in Table~\ref{tab4} demonstrate that near cancelation of the $^1S_0$ and
$^3P_0$ state polarizabilities in divalent B$^+$, Al$^+$, In$^+$ ions of group IIIB \cite{SafKozCla11} continues for much heavier Tl$^+$, leading to
anomalously small BBR shift for this system.  This calculation demonstrates that the BBR contribution to the fractional frequency uncertainty of the
Tl$^+$ frequency standard at $300~K$  is $1\times10^{-18}$. We find that Tl$^+$ has the smallest fractional BBR shift among all present or proposed
frequency
standards with the exception of Al$^+$.\\
\section*{ACKNOWLEDGEMENTS}
We thank Hugh Klein for bring our attention to the problem of BBR shift in Tl$^+$ frequency standard.  This work was supported in part by US NSF
Grants No.\ PHY-1068699 and No.\ PHY-0758088. The work of MGK was supported in part by RFBR grant No.\ 11-02-00943.

\end{document}